\documentclass{PoS}
\usepackage{slashed}
\title{Universal Correlations in Pion-less EFT with the Resonating Group Model: Three and Four Nucleons}
\ShortTitle{Universality in Nuclei with $A\le 6$}
\author{\speaker{Johannes Kirscher}, Harald W.~Grie\ss hammer, Deepshikha Shukla\\
        Center for Nuclear Studies, Department of Physics, The George Washington University, Washington, DC  20052, USA\\
        E-mail: \email{kirscher@gwmail.gwu.edu}}
\author{Hartmut M.~Hofmann\\
        Institut f\"ur Theoretische Physik III, Department f\"ur Physik, Universit\"at Erlangen-N\"urnberg, D-91058 Erlangen, Germany}
\abstract{A systematic connection between QCD and nuclear few- and many-body
properties in the form of the Effective Field Theory ``without pions'' is
applied to $A\le 6$ nuclei to determine its range of applicability. We present results at next-to-leading
order for the Tjon correlation and for a correlation between the singlet S-wave
$^3$He-neutron scattering length and the triton binding energy.
In the $A=6$ sector we performed leading order calculations
for the binding energy and the charge and matter radii of the halo nucleus $^6$He. Also at leading order,
the doublet S-wave 4-He-neutron phase shifts are compared with R-matrix data. These analysis provide evidence for a sufficiently fast convergence
of the effective field theory, in particular, our results in $A\le 4$ predict an expansion parameter of
about $\frac{1}{3}$, and they converge to data within the predicted uncertainty band at this order.
A properly adjusted three-body contact force which we include together with the
Coulomb interaction in all calculations is found to correctly renormalize the pion-less theory at leading- and next-to-leading order, \textit{i.e.},
the power counting does not require four-body forces at the respective order.
}
\FullConference{6th International Workshop on Chiral Dynamics, CD09\\
		July 6-10, 2009\\
		Bern, Switzerland}
\begin{document}
\section{Introduction}
In principle, effective field theories (EFT) provide a framework in which properties of nuclei can be derived
from the underlying theory of QCD. In contrast to the elegance and success of Chiral Perturbation theory in the mesonic sector,
its extention, by including baryonic degrees of freedom, called
Chiral Effective Field Theory, suffers practical and systematic problems if it is used to expand
multi-baryon amplitudes~\cite{xpt-pc-probl}. There are no such inconsistencies in the power counting of
the Effective Field Theory ``without pions'' (EFT$_\slashed{\pi}$, see \textit{e.g.}~\cite{piless-rev}).
%similar to the ones encountered in multi-baryon $\chi$PT stemming from pion exchanges, have been found.
Furthermore, this EFT has a simple vertex structure with nucleon-nucleon (NN) contact terms only, which makes calculations in
larger nuclei more practical.\\
EFT$_\slashed{\pi}$ systematically expands observables in powers of the typical low-momentum scale
$p_\textrm{{\tiny typ}}$ of the process considered, measured in units of the scale $\Lambda_{b}$
at which the pion can be resolved as a dynamical degree of freedom, set by the pion mass $m_\pi$.
While this expansion has been shown to converge sufficiently fast for many reactions in the two- and three-nucleon
sector with and without electro-weak currents (see \textit{e.g.}~\cite{piless-rev}), its usefulness for systems with $A\ge 4$ has been under question.
The question whether or not an observable is apt to be expanded within EFT$_\slashed{\pi}$ is relatively
easy to answer for two nucleons where $p_\textrm{{\tiny typ}}$ is closely related to, \textit{e.g.}, the relative momentum
in neutron-proton scattering, or the binding momentum of the deuteron. In contrast, the $^4$He binding energy
$B(^4\textrm{He})=28.5$~MeV, and its typical size of about $1.6$~fm is comparable with the approximate range of the
one-pion exchange. A systematic expansion in the four-nucleon system could therefore converge very slowly,
at best, if the na\"ive estimate for $p_\textrm{{\tiny typ}}$ turns out to be appropriate. The current understanding of
effective field theories does not provide rigorous arguments at which values, \textit{e.g.}, of $A$ or of the density inside the nucleus,
EFT$_\slashed{\pi}$ fails. Explicit calculations are required!\\
To demonstrate the usefulness of EFT$_\slashed{\pi}$ applied to the 4-nucleon bound- and scattering
system, we analyzed the Tjon correlation and a correlation between the singlet S-wave
$^3$He-neutron scattering length $a_0$($^3$He-n) and the triton binding energy $B(t)$.
The results are taken from the more detailed analysis in~\cite{wannabe} (an extensive list of references can be found there).
With the systematic and predictive power thus demonstrated, the status of our investigation in an $A=5$ scattering system
and the halo system $^6$He is shown. Specifically, we present leading order (LO) results for the doublet S-wave phase shift for $^4$He-neutron scattering,
in addition to binding energy, charge- and matter radii results for $^6$He. Before that, we present the specific form of the
next-to-leading order (NLO) EFT$_\slashed{\pi}$ NN interaction and the few-body technique of the
Refined Resonating Group Model (RRGM~\cite{rrgm}) used to solve the many-body Schr\"odinger equation.
\section{Pion-less Theory of the Nuclear Interaction}
The interaction part of the Lagrangean of EFT$_\slashed{\pi}$ at NLO (see \textit{e.g.}~\cite{piless-l}) consists of nine four-nucleon contact terms and one
six-nucleon vertex. A spin dependend term, one spin independent one, and the six-nucleon term are momentum
independent and comprise the LO part of the potential which is to be iterated to all orders because of the unnaturally
large scattering lengths $a_{s,t}$ in the NN system. At NLO vanishing P-wave amplitudes result in dependencies amongst
the seven new coupling strengths and reduce their number to only three independent NLO parameters.\\
We solve the Schr\"odinger equation in coordinate space with the following operator basis:
\begin{small}
\begin{eqnarray}\label{eq.pot-coord}
V^{NLO}_{\textrm{EFT}_\slashed{\pi}}(\vec{r})&=&
e^{-\frac{\Lambda^2}{4}\vec{r}^2}\left(A_1+A_2\vec{\sigma}_1\cdot\vec{\sigma}_2\right)+\left(A_3+A_4\vec{\sigma}_1\cdot\vec{\sigma}_2\right)\Big\lbrace
e^{-\frac{\Lambda^2}{4}\vec{r}^2},\vec{\nabla}^2\Big\rbrace+\nonumber\\
&&e^{-\frac{\Lambda^2}{4}\vec{r}^2}\left(A_5+A_6\vec{\sigma}_1\cdot\vec{\sigma}_2\right)\vec{r}^2+e^{-\frac{\Lambda^2}{4}\vec{r}^2}A_7\vec{L}\cdot\vec{S}+
e^{-\frac{\Lambda^2}{4}\vec{r}^2}A_8\left(\vec{\sigma}_1\cdot\vec{r}\vec{\sigma}_2\cdot\vec{r}-\frac{1}{3}\vec{r}^2\vec{\sigma}_1\cdot\vec{\sigma}_2\right)-\nonumber\\
&&A_9\Big\lbrace e^{-\frac{\Lambda^2}{4}\vec{r}^2},\Big[\big[\partial^r\otimes\partial^s\big]^{2}\otimes\big[\sigma_1^p\otimes\sigma_2^q\big]^{2}\Big]^{00}\Big\rbrace+
A_{3NF}\,e^{-\frac{\Lambda^2}{4}\vec{r_{12}}^2}e^{-\frac{\Lambda^2}{4}\vec{r_{23}}^2}\,\vec{\tau}_1\cdot\vec{\tau}_2\;.\nonumber\\
\end{eqnarray}
\end{small}
By changing the renormalization scheme in two distinct ways, we test the consistency of the power counting and get an
estimate of higher order effects, \textit{i.e.}, the convergence rate of the theory. The two methods are:
\begin{enumerate}
 \item A variation of the parameter $\Lambda$ of the Gaussian regulator in the interval $[400;700]$~MeV.
 \item A variation of the data used to fit the low-energy constants (LEC) $A_i$: We did not implement the
           dependencies between the $A_i$'s but fitted nine coupling strengths independently to allow for non-zero P-wave phase shifts
           up to $0.1$\% of the Nijmegen values.
\end{enumerate}
Both represent a modification of the unresolved short distance structure of the nucleons and should affect
low-energy observables only within the uncertainty range of the considered order. Nine NLO potentials, corresponding
to nine different sets of $A_i$'s, were fitted to the deuteron binding energy $B(d)$ and to the neutron-proton (np) phase
shifts $\delta(^{(1,3)}S_{0,1})\,,\epsilon_1(^3S_1-^3D_1)\,$ below $0.3$~MeV from~\cite{nij-pwa}. Each LEC set was determined either with a different cutoff
$\Lambda$ or different values for the P-wave phase shifts. The three-body input to fix $A_{3NF}$ was $B(t)$, and the
LO parameters were adjusted to the singlet and triplet np scattering lengths.\\
Corrections to the non-relativistic kinetic energy term are of higher order. Hence, amplitudes are calculated by first iterating
the LO part of the potential to all orders by solving, \textit{e.g.}, the Schr\"odinger equation, and a consecutive perturbative treatment
of the NLO operators in distorted wave Born approximation. However, we iterate also the NLO vertices and conclude from the
results that higher order diagrams which are included in such an approach yield only negligible contributions.
\section{Resonating Group Model}
To reach the goals stated in the introduction, we need a method to solve the few-body bound- and scattering problem. We employ
a refined version of the Resonating Group Model which yields a variational solution to the Schr\"odinger equation in coordinate
representation (see \textit{e.g.}~\cite{rrgm}). The variational space for a N-body bound state is spanned by a superposition of antisymmetric ($\mathcal{A}$) states
of the form:
\begin{small}
\begin{equation}\label{eq.bsbv}
\psi\left(\vec{\rho}_m,\vec{s}_m\right)=
\mathcal{A}\left\lbrace\sum_{d,i,j}c_{dij}\Bigg[\Big[\prod_{k=1}^{N-1}e^{-\gamma_{dk}\vec{\rho}_k^2}
\mathcal{Y}_{l_{ki}}\left(\vec{\rho}_k\right)\Big]^{L_i}\otimes\Xi^{S_j}\Bigg]^J\right\rbrace\;\;\;\;.
\end{equation}
\end{small}
The index $(i) j$ labels different (orbital) spin angular momentum coupling schemes and thus resembles the idea of the resonating groups,
\textit{i.e.}, a hierarchy of all possible groupings of neutrons and protons within a nucleus which allows a simplification of the wave
function ansatz by discarding those states corresponding to groupings that are relatively short lived. This simplification, irrelevant thanks to modern computer technology
for the three- and four-body observables for which we included all possible groupings, becomes an efficient
% however non-systematic
tool that makes $A>5$ body calculations feasible. For instance, a single fragmentation with a $^4$He core and two attached neutrons
was found to yield more than $95$\% of the assumed converged observables calculated here for $^6$He.\\
The dimension of the model space is further reduced by including only orbital angular momenta $l_{ki}\le 2$ on the Jacobi coordinates
$\vec{\rho}$. In addition, certain couplings even with $l=1$, \textit{e.g.}, for the coordinate between the $^4$He core and the center of mass
of the two skin neutrons and the relative coordinate between those neutrons, an explicit calculation showed the contribution to $B(^6\textrm{He})$
to be negligible.\\
Finally, the number and magnitudes of the Gaussian width parameters $\gamma_d$ had to be tailored to the various potentials to
allow for enough flexibility of the basis in the range where the potential is non-zero. This interval varies with the cutoff $\Lambda$.
The numerical instabilities due to the non-orthogonality of the basis were monitored to ensure also a numerically independent set of basis vectors.
\section{Results}
\subsection{Four nucleons: bound state}
\begin{figure}
\begin{minipage}[t]{.5\textwidth}
\includegraphics[width=\textwidth]{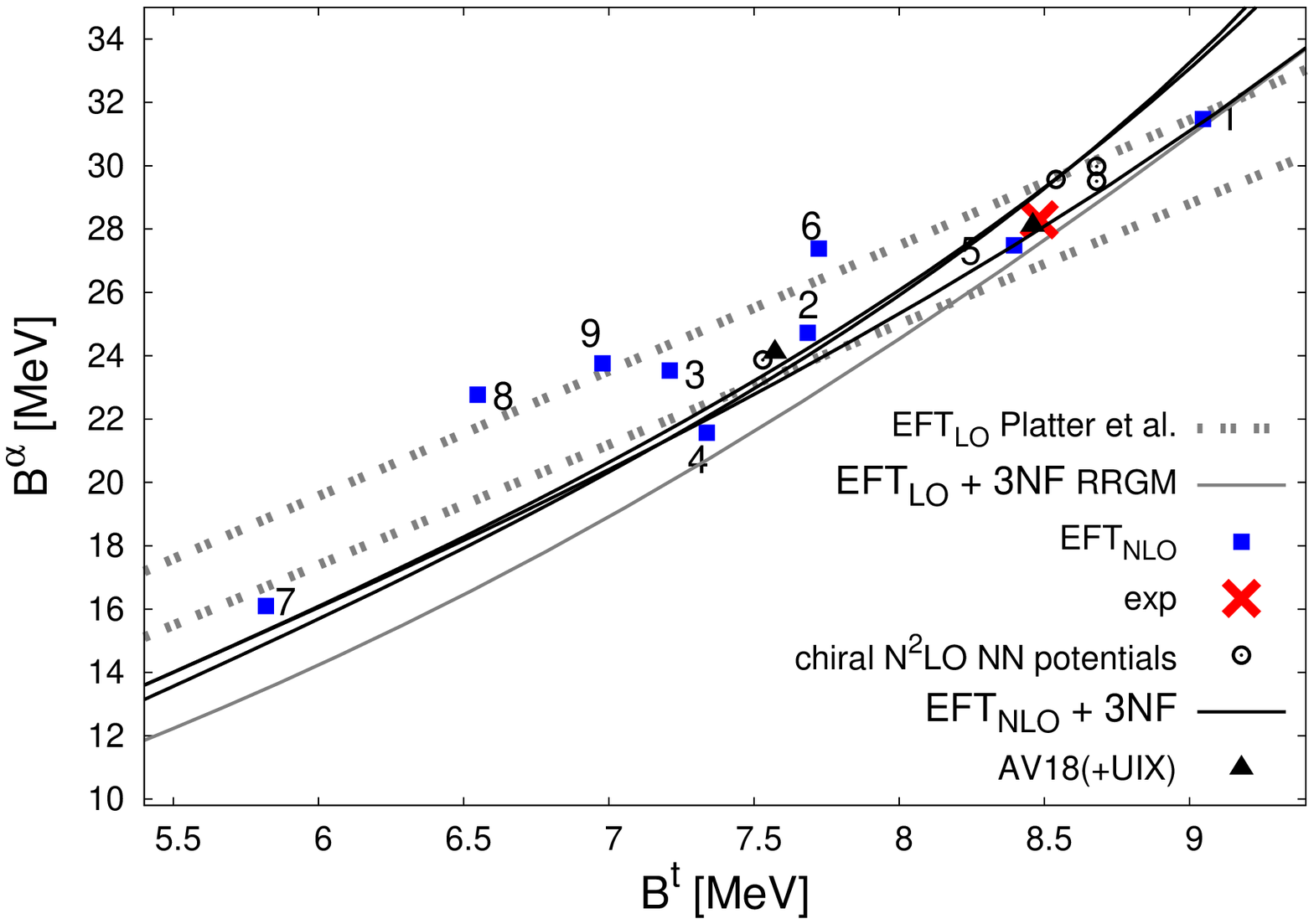}
\end{minipage}\hfil
\begin{minipage}[t]{.5\textwidth}
\includegraphics[width=\textwidth]{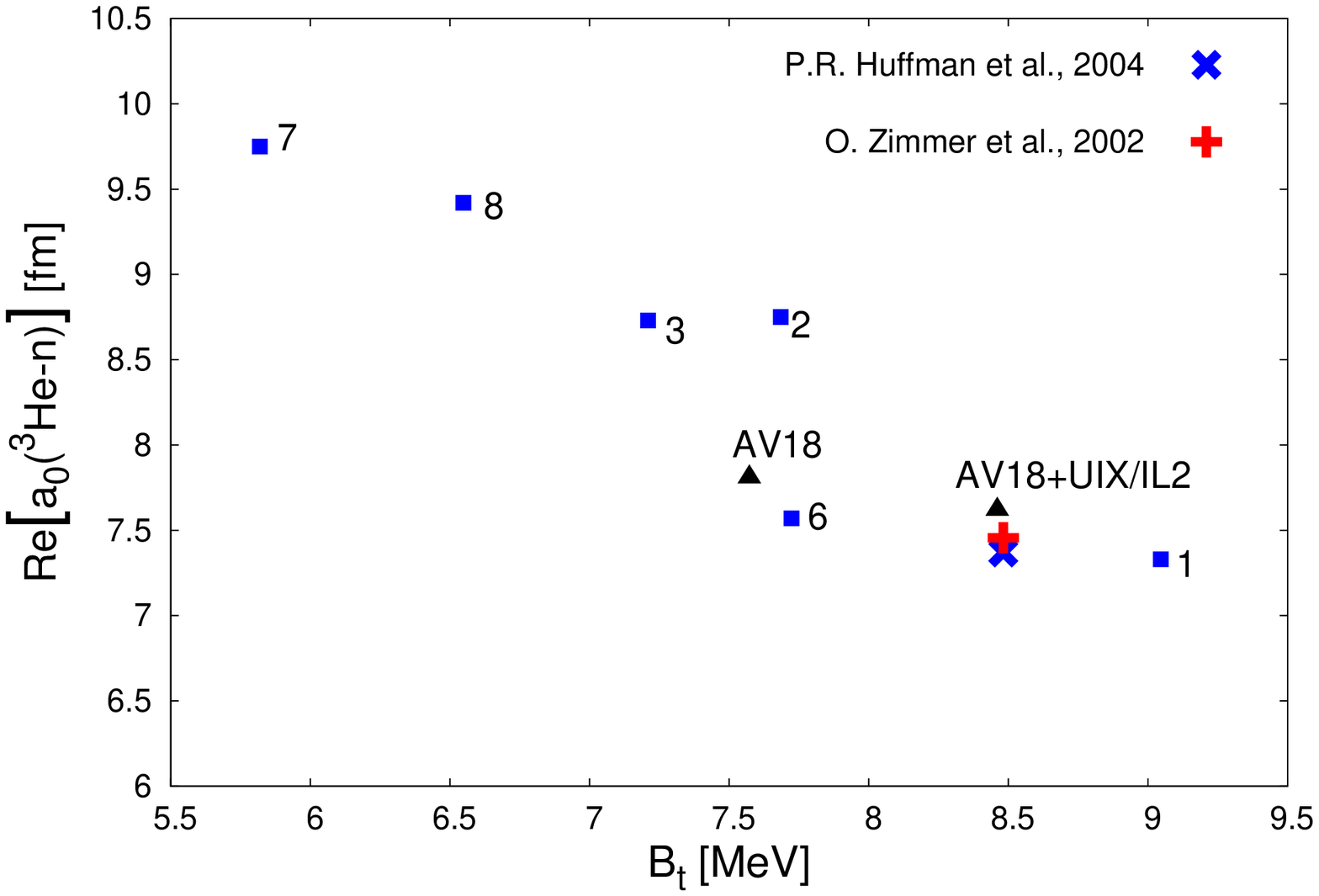}
\end{minipage}
\caption{\label{fig.tjon}{\small \textbf{(left)} Correlation between the triton and $^4$He
    binding energies (Tjon line). The squares are results from the nine NLO
    EFT$_\slashed{\pi}$ NN potentials. The solid (gray) black lines represent
    a smooth variation of $A_{3NF}$ for a specific (LO) NLO potential.
    The upper (lower) dashed line was calculated at LO in~\cite{platter-tjon} with LECs fitted to $a_{s,t}$ ($B(d),a_s$).
    The values for AV18(+UIX) are reported in~\cite{mythesis}, and those using a chiral expansion to NLO and N$^2$LO
    in~\cite{xpt-epelbaum}.\newline
\textbf{(right)} The correlation between the triton binding energy
    and the real part of the spin singlet, S-wave scattering length
    $a_0(^3\textrm{{\small He-n}})$ for elastic $^3$He-n scattering. The dots
    represent the values of the NLO EFT$_\slashed{\pi}$ potentials. The values for
    AV18(+UIX/IL2)~\cite{il2-pot} are results of RRGM calculations reported in~\cite{mythesis}}.}
\end{figure}

EFT$_\slashed{\pi}$ should appropriately describe the $^4$He bound state to make its application
to larger systems, where this state plays an important role as a target or core, a worthwile endeavor.
We analyzed the dependence of $B(^4\textrm{He})$ on $B(t)$ using nine NLO potentials with an unconstrained
$A_{3NF}$ parameter, \textit{i.e.}, the two-body sector is renormalized properly but for $A>2$ one expects
deviations from data analogous to older model calculations with NN-forces only. With an EFT, this dependence,
known as Tjon-line, becomes a band whose width is a measure for the accuracy of the expansion. A change
in the renormalization scheme, in our case a change in the cutoff $\Lambda$ and the P-wave phase shifts,
should yield results for the binding energy tuple within an uncertainty range set by $\left(p_\textrm{{\tiny typ}}/m_\pi\right)^n$
around the datum in an N$^n$LO calculation.\\
In the left panel of Fig.~\ref{fig.tjon}, the results are compared to the LO band calculated by Platter \textit{et al.}~\cite{platter-tjon} and with our
LO potential. The spread of the NLO values is not in conflict with EFT$_\slashed{\pi}$ which predicts a $10$\%
uncertainty at NLO. The upper (lower) boundary of the LO band was obtained by choosing different NN
observables, $a_{s,t}(B(d),a_s)$, to fit the LO LECs, and it was already pointed out in~\cite{platter-tjon} that this gives only a
crude estimate of higher order effects. The difference between the $^4$He binding energies obtained
with the two potentials \#2 and \#6, which yield almost identical triton binding energies, serves as an estimate of
the theoretical uncertainty. With this estimate and the points from the EFT$_\slashed{\pi}$ potentials
in Fig.~\ref{fig.tjon} (left), we deduce a NLO correlation band of width of about $5~$MeV centered around $28~$MeV at the experimental $B(t)$.
The resultant prediction of
\begin{equation}\label{eq.b4he-nlo}
B(^4\textrm{He})^{\textrm{{\tiny NLO}}}=(28\pm 2.5)~\textrm{MeV}
\end{equation}
is consistent with the expected NLO uncertainty of about $10$\% and with experiment.
The results of the AV18(+UIX) models lie within the proposed band as it is expected
of all interaction models of at least NLO.\\
Each of the nine squares in Fig.~\ref{fig.tjon} corresponds to a set of $A_i$'s with $A_{3NF}=0$. Focusing on point
\#1, the solid lines result from a smooth variation of $A_{3NF}$ with the other LECs of the respective NLO and LO potentials held fixed. Both lines pass the
datum well within the uncertainty bounds. The absolute value of $A_{3NF}$ depends on the renormalization
scheme chosen to obtain the two-body LECs. Our results also show at NLO that the two-nucleon force
can be chosen such that $A_{3NF}=0$. In general a non-zero three-nucleon counter term is mandatory
for a correct renormalization of the triton. This term corrects an unphysical dependence of A > 2 low-energy
observables on details of short-distance NN interactions, while low-energy NN predictions are unaffected.
A three-nucleon interaction is therefore necessary to guarantee renormalization scheme independence of A > 2 low-energy observables.
Once $A_{3NF}$ is fitted to appropriate data, here $B(t)$, all remaining dependence on the renormalization parameters, here $\Lambda$ and input data, in $B(^4\textrm{He})$
is a less than 10\% higher order effect. This is pictured in Fig.~\ref{fig.tjon} for two NLO and one LO potential,
and we present it as evidence that no additional four-body parameter is needed at NLO,
and hence the four-nucleon bound state is a universal consequence of the three-body system. Furthermore, the LO
line (light gray), deviates more from the center of the correlation band, which is a confirmation of the order-by-order
improvement of the accuracy of the EFT$_\slashed{\pi}$ calculation. For a detailed analysis
we refer to a forthcoming report~\cite{3nf-forth}.
\subsection{Four nucleons: scattering}
In principle, all low-energy observables should be correlated with the triton binding energy via one parameter at NLO if the
assertion that four-nucleon forces are at least N$^{2}$LO is true. Also, the uncertainty in the prediction of other low-energy
observables due to the truncation of the Lagrangean should be comparable to the one identified in the previous subsection.
We test both hypotheses by looking at the real part of the S-wave spin singlet scattering length $a_0(^3\textrm{{\small He-n}})$ for elastic
$^3$He-neutron scattering. Its dependence on $B(t)$ is shown for six NLO EFT$_\slashed{\pi}$ potentials including the
Coulomb interaction in the right panel of Fig.~\ref{fig.tjon}.\\
We find the na\"ively expected decrease of $\textrm{Re}\lbrace a_0(^3\textrm{{\small He-n}})\rbrace$ with increasing
triton binding energy which maps out a band including the datum. Consistent with the analysis of the Tjon-line, the
error estimate we deduce from the change in $a_0(^3\textrm{{\small He-n}})$ observed between the potentials  \#2 and \#6
is about $10$\%. Again, the assumption of the center of the band at the experimental $B(t)$
follows from the values in Fig.~\ref{fig.tjon} (right). Thus, EFT$_\slashed{\pi}$ reports
\begin{equation}\label{eq.a0-value-nlo}
\textrm{Re}\lbrace a_0(^3\textrm{{\small He-n}})\rbrace^\textrm{{\tiny NLO}}=\left(7.5\pm 0.6\right)~\textrm{fm}\;\;\;.
\end{equation}
The NN model AV18 yields a value within the error band. Adding the UIX/IL2 three-body
force (for details about those calculations see~\cite{mythesis,hmh-4he}) moves this point into the $10$\% NLO
uncertainty range around the datum. This example of a low-energy scattering observable supports our previous conclusions:
Every potential with the correct NN low-energy phase shifts and appropriately tuned three-body force, \textit{e.g.},
to give the correct triton binding energy, should not only predict the correct $B(^4\textrm{He})$ but also the experimental
$a_0(^3\textrm{{\small He-n}})$ within a NLO error range of about $10$\%. No additional four-nucleon parameter
is therefore necessary.
\subsection{Five and six nucleons}
\begin{figure}
\begin{minipage}[t]{.5\textwidth}
\includegraphics[width=\textwidth]{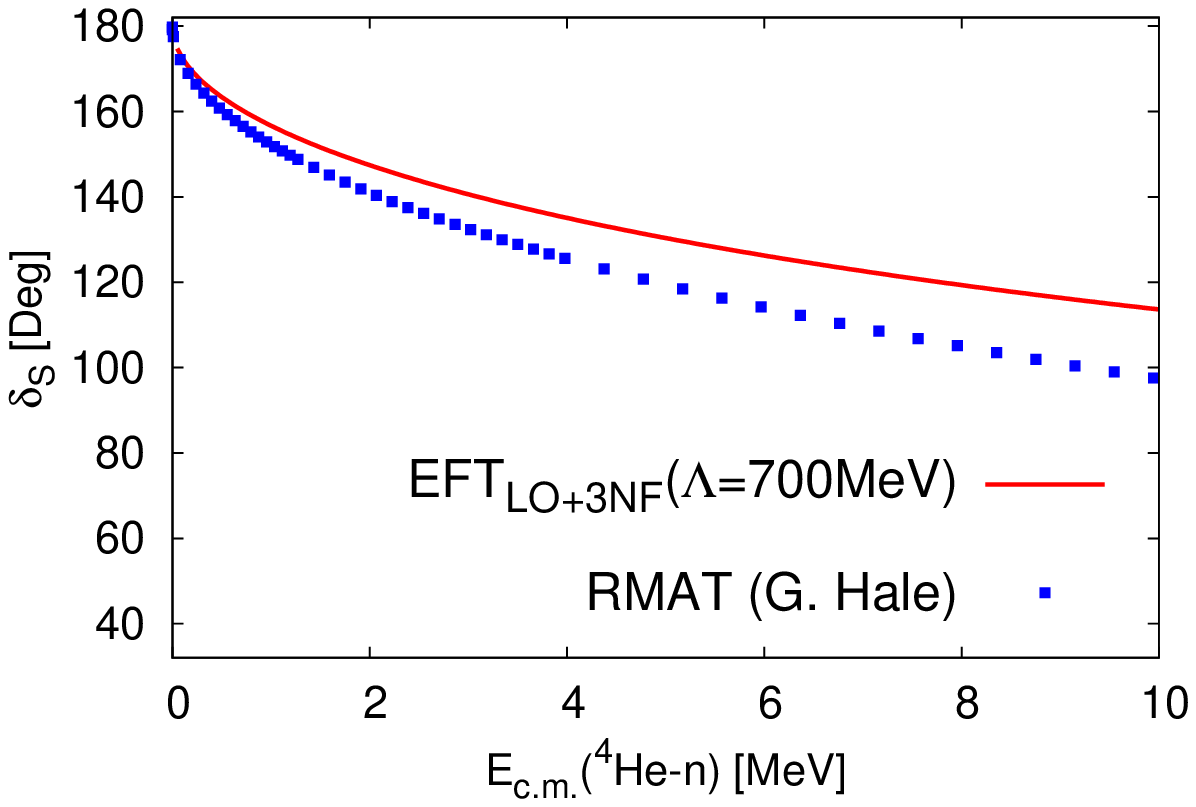}
\end{minipage}\hfil
\begin{minipage}[t]{.5\textwidth}
\includegraphics[width=\textwidth]{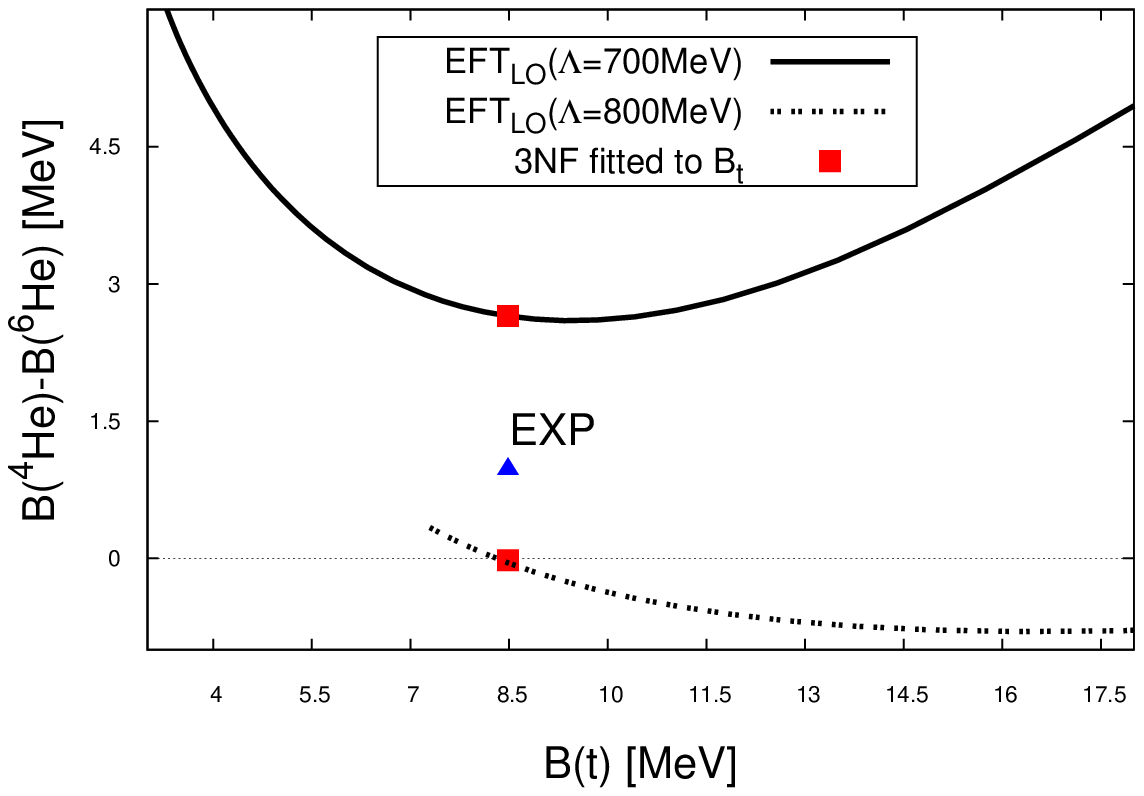}
\end{minipage}
\caption{\label{fig.5he6he}{\small \textbf{(left)} S-wave phase shift in the $\frac{1}{2}^+-$channel for $^4$He-n scattering. Compared
are the results from a LO EFT$_\slashed{\pi}$ calculation (red, solid line) to an R-matrix analysis (blue squares,~\cite{hale-rmat}).\newline
\textbf{(right)} Splitting between the $^6$He and $^4$He (lowest breakup threshold) binding energies. The thick (dashed) solid line was
obtained with a LO EFT$_\slashed{\pi}$ potential with the two-nucleon force parameters held fixed while smoothly changing
the three-body force. The set of RRGM parameters, the model space, was only optimized for the potential with $\Lambda=700$~MeV}.}
\end{figure}

The results presented in this subsection should give an outlook and serve as a demonstration of the feasibility of
calculations in heavier systems with three-body- and Coulomb forces. Here, the $^4$He-neutron $\left(\frac{1}{2}^+\right)$-phase shifts
were calculated as an intermediate step before looking at the $^6$He ground state. A bound state in this $A=5$ channel
would indicate a possible failure of EFT$_\slashed{\pi}$ within its current ordering scheme. From the LO results shown in
Fig.~\ref{fig.5he6he} (left) obtained with a cutoff $\Lambda=700$~MeV and $A_{1,2,3NF}$ fitted to $a_{s,t},B(t)$, we conclude, preliminary,
that no such state exists, and hence a modification of the power counting at LO is unnecessary.\\
With this potential we calculated two observables in the $^6$He system which we take as signatures of a halo structure.
First, a relatively small charge radius $\langle \vec{r}^2\rangle_{\textrm{{\tiny ch}}}^{1/2}$ compared to the matter radius $\langle \vec{r}^2\rangle_{\textrm{{\tiny m}}}^{1/2}$,
and second, a shallow bound state with respect to the $^4$He-nn breakup threshold.
The qualitative behavior of the two radii we calculated,
\begin{equation}
\langle \vec{r}^2\rangle_{\textrm{{\tiny ch}}}^{1/2}= 3.261~\textrm{fm}\;\;\;,\;\;\;\langle \vec{r}^2\rangle_{\textrm{{\tiny m}}}^{1/2}=5.678~\textrm{fm}\;\;\;,
\end{equation}
hints at a halo structure. We attribute the quantitative discrepancy to data in part to the small model space used for the $^4$He core in which it
is only bound by $B(^4\textrm{He})\approx 20$~MeV.
We expect a refinement of the variational basis to yield values consistent with EFT error margins. However, these margins, \textit{i.e.}, the size of the expansion
parameter of  EFT$_\slashed{\pi}$ in $A=6$, have yet to be determined from the LO to NLO convergence and cannot be estimated \emph{a priori} as the example of $^4$He shows.
In Fig.~\ref{fig.5he6he} (right) the dependence of $\left(B(^4\textrm{He})-B(^6\textrm{He})\right)$ on $B(t)$ is plotted for a varying $A_{3NF}$. For
$\Lambda=700$~MeV, the curve has a minimum close to the three-fermion system which corresponds to the physical triton.
This means, that the halo structure is a peculiar property of nuclei, which is less pronounced in systems with
the same two-body structure but different trimer properties. The halo character of the 6-body state only follows for a certain range
of the binding energy of the three-body bound state. It is not a universal consequence of the unnaturally large two-body scattering lengths alone.\\
To confirm this finding, the aforementioned $^4$He model space has to be enlarged to reach convergence in $B(^4\textrm{He})$. A NLO calculation
will yield an estimate of the expansion parameter of EFT$_\slashed{\pi}$ in $A=6$. At this stage of our analysis it is an open question whether
or not $^6$He is bound, a possible, relatively slow EFT convergence would also allow for an unbound nucleus.
The value of the expansion parameter will ultimately tell if the existence of a bound $^6$He is a higher than NLO effect or not.
Disregarding numerical issues related to the RRGM for the moment, the dependence for $\Lambda=800$~MeV
shows this possibility of an unbound system, \textit{i.e.}, with the change in the renormalization scheme, the shallow halo state ($\Lambda=700~$MeV) disappears ($\Lambda=800$~MeV).
However, we attribute this observation primarily to the RRGM model space used, specifically the set of
Gaussian parameters (see Eq.~(\ref{eq.bsbv})) on the coordinates for the skin neutrons which were optimized for $\Lambda=700$~MeV.
From this analysis we deduce the following roadmap for $^{(5,6)}$He calculations to solidify our assertions.
\begin{enumerate}
 \item Check renormalization scheme independence by varying the cutoff ($\Lambda\in[300;800]$~MeV) and by different input data (different NN P-waves,
and $A_{3NF}$ fitted to the charge radius of the triton instead of $B(t)$ or the respective values of $^4$He).
\item For each potential thus fitted, the variational space has to be refined until full convergence is established.
\item A NLO calculation with a \textit{clean}, perturbative treatment of momentum dependent vertices has to be performed to
demonstrate the convergence of the EFT expansion.%, and by that its superiority over phenomenology.
\item Observables of subsystems have to be monitored carefully for consistency with the predicted theoretical uncertainty at the given order.
For instance, the model spaces used for the $^4$He cores do not meet this criterium and have to be refined as explained above.
\end{enumerate}

\section{Conclusions}
We presented an analysis of two universal correlations, the dependencies of the $^4$He binding energy and a $^3$He-neutron scattering
length on the triton binding energy, with the nuclear Effective Field Theory ``without pions''. The calculations include a three-body force
and the Coulomb interaction, and report the expected order-by-order improvement consistent with an expansion parameter
$p_\textrm{{\tiny typ}}/\Lambda_{\textrm{{\tiny b}}}\approx\frac{1}{3}$. By that we demonstrated that a systematically improveable description of the
$^4$He-nucleus is possible with EFT$_\slashed{\pi}$. In particular, we find no evidence that a four-body contact interaction is required to
renormalize this system at next-to-leading order. The tetramer is a universal consequence of the trimer.\\
We also presented preliminary results of an analysis in the five-nucleon scattering system, and of halo signatures of the $^6$He system. The former
reproduces the R-matrix data within leading order error margins while the latter yields the qualitative halo features but so far does not
allow for final conclusions before certain numerical issues are carefully investigated. At this point of our analysis, the halo structure of $^6$He
is a universal consequence of two- and three-nucleon properties, while unnaturally large two-nucleon scattering lengths alone are not
sufficient for a peculiar system like that.\\
In a next step, we have to confirm the findings for the $A>4$ systems at leading order by following the roadmap laid out in the previous section.
Then, the theory will be used for predictions in nuclear systems with $A\le 8$, including (un)stable Lithium and Beryllium states. We also
aim towards an application to bosonic systems of atomic gases where recent measurements~\cite{cold-exp} allow for an \emph{observation} of the change of the
renormalization scheme. As a prof of principle, a matching of EFT$_\slashed{\pi}$ to a cluster effective field theory (\textit{e.g.}~\cite{halo-eft}) would demonstrate this bottom-up
approach of the construction of an EFT.
\end{document}